\documentclass[aps,prl,reprint,superscriptaddress]{revtex4-2}

\usepackage{graphicx}
\usepackage{dcolumn}
\usepackage{bm}
\usepackage{color}
\usepackage{xcolor}
\usepackage{epstopdf}
\usepackage{gensymb}
\usepackage{ulem}
\usepackage{float}
\usepackage{xr}
\usepackage{sidecap}
\usepackage{mathrsfs}
\usepackage{amsmath}
\usepackage{amsthm}
\usepackage{enumerate}
\usepackage{soul}
\usepackage{amssymb}
\usepackage{units}
\usepackage{afterpage}
\usepackage{lineno,hyperref}
\begin{document}
\preprint{APS/123-QED}

\title{
Fractal Patterns in the Parameter Space of Bi-stable Duffing Oscillator\\}

\author{Md Nahid Hasan}
\affiliation{Department of Mechanical Engineering, University of Utah, Salt Lake City, UT 84112, USA}

\author{Taylor E. Greenwood}%
\affiliation{Department of Mechanical Engineering, University of Utah, Salt Lake City, UT 84112, USA}

\author{Robert G. Parker}%
\affiliation{Department of Mechanical Engineering, University of Utah, Salt Lake City, UT 84112, USA}

\author{Yong Lin Kong}%
\affiliation{Department of Mechanical Engineering, University of Utah, Salt Lake City, UT 84112, USA}
\thanks{yong.kong@utah.edu}

\author{Pai Wang}%
\affiliation{Department of Mechanical Engineering, University of Utah, Salt Lake City, UT 84112, USA}
\thanks{pai.wang@utah.edu}

\begin{abstract}
We study the dissipative bi-stable Duffing oscillator with equal energy wells and observe fractal patterns in the parameter space of driving frequency, forcing amplitude, and damping ratio. Our numerical investigation reveals the Hausdorff fractal dimension of the boundaries that separate the oscillator’s intra-well and inter-well behaviors. Furthermore, we categorize the inter-well behaviors as three steady-state types: switching, reverting, and vacillating. While fractal patterns in the phase space are well-known and heavily studied, our results point to a new research direction about fractal patterns in the parameter space. Another implication of this study is that the vibration of a continuous bi-stable system modeled using a single-mode approximation also manifests fractal patterns in the parameter space. In addition, our findings can guide the design of next-generation bi-stable and multi-stable mechanical metamaterials. 
\end{abstract}
\maketitle
\begin{figure}[H]
\centering
\includegraphics[scale=0.45]{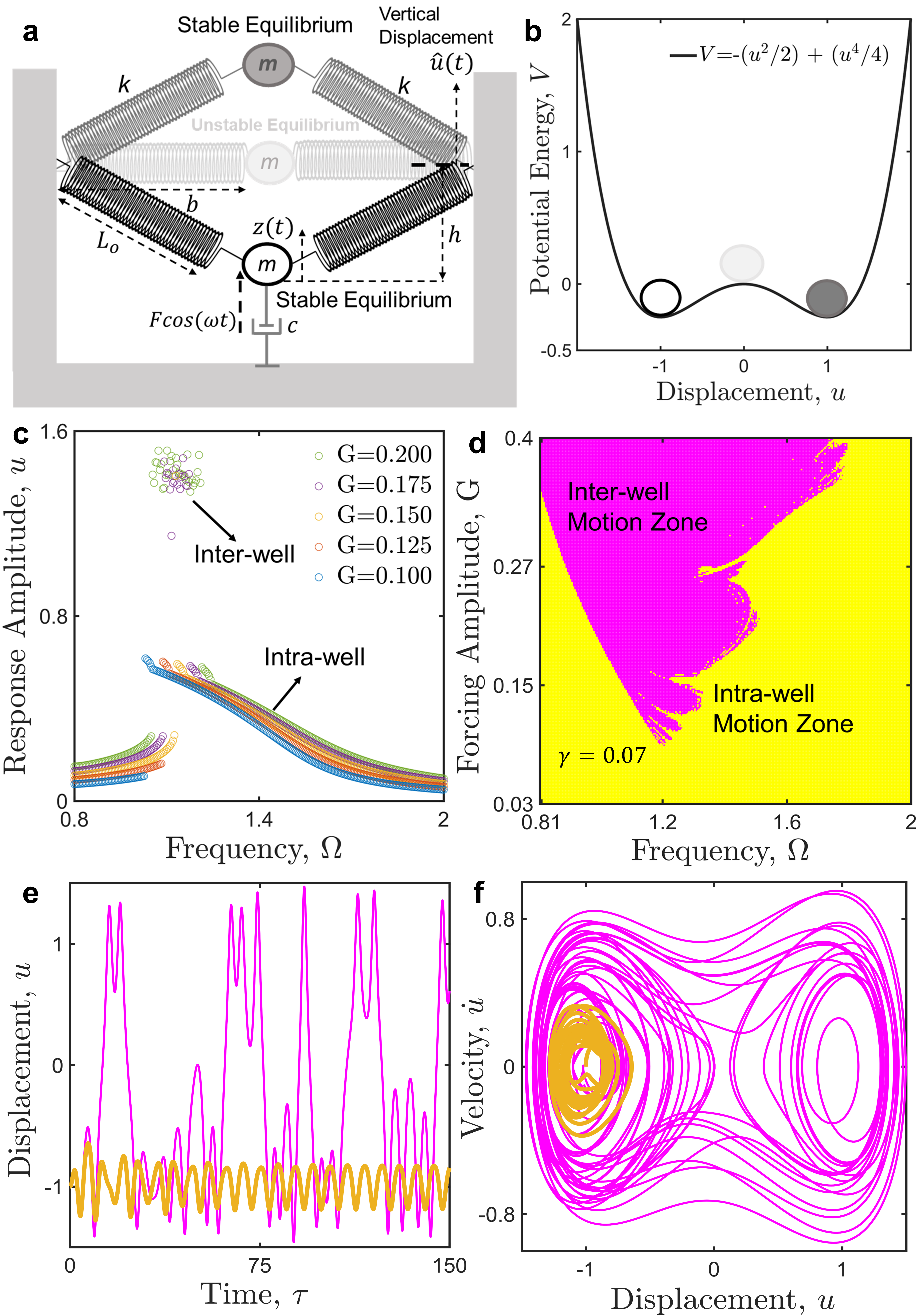}
\caption{\label{fig:F1} 
(a) Physical setup that behaves as a dissipative bi-stable Duffing oscillator. Here, the zero-force spring length $L_o>b$ makes the central configuration (the gray-shaded state in the middle) unstable. (b) Double-well potential with an energy barrier at the center with $\mu=1$. (c) Five frequency response curves across values of forcing amplitude $G$. (d) Forcing amplitude-frequency ($G$\,vs.\,$\Omega$) parameter space where magenta and yellow zones represent inter-well and intra-well behavior regions, respectively. (e) Time response curves show the inter-well (magenta) and intra-well (yellow) behaviors. (f) Phase spaces of inter-well (magenta) and intra-well (yellow) behaviors.}
\end{figure}

\textit{Introduction:} The dissipative bi-stable Duffing oscillator is a well-known dynamical system with applications in many fields, such as shape morphing~\cite{cao2021bistable,arrieta2013dynamic,diaconu2008concepts,bilgen2013dynamic,mattioni2008analysis,carrella2008numerical,carrella2008static,schenk2014review, wu2022programming, pontecorvo2013bistable}, energy harvesting~\cite{arrieta2010piezoelectric,harne2013review,senba2010two,stanton2012harmonic,emam2015review,chen2021curve, huguet2019parametric,qian2022bio}, soft robotics~\cite{chen2018harnessing,rothemund2018soft,chi2022bistable}, MEMS devices~\cite{matoba1994bistable,casals2008snap, hussein2019design, qiu2004curved,wang1998chaos,shaw2017nonlinearity}, energy absorption~\cite{montalbano2020design,giri2021controlled,sun2019snap}, and drug delivery~\cite{wagner1996micromachined}. In addition, with recent advances in the multi-stable metamaterial~\cite{zhang2022bioinspired,bilal2017bistable,intrigila2022fabrication,chen2014bistability, shan2015multistable,mao2022modular,haghpanah2016multistable,wang20223d,van20214d,boley2019shape,shi2021programmable,hasan2023optimization,greenwood2023advanced}, and advanced functional systems~\cite{alapan2020reprogrammable,kim2018printing,bai2022dynamically,tang2022wireless}, bi-stable system dynamics can be introduced as a mechanism for folding~\cite{sadeghi2020dynamic,fang2017dynamics} and reprogramming~\cite{eichelberg2022metamaterials}. As illustrated in Fig.\,\ref{fig:F1}(a), one physical realization of the dissipative bi-stability consists of two linear elastic connections of equal spring constants, one linear damping, and a point mass. The system has two stable equilibria and one unstable equilibrium. The generalized governing equation is:
\begin{equation}\label{eq:1}
\begin{split}
m \frac{d^2 \hat{u}}{d t^2}+ c\frac{d \hat{u}}{d t} - k_1 \hat{u}+ k_3\hat{u}^3=F\cos(\omega t),
\end{split}
\end{equation}
which contains a cubic non-linearity term emerging from the geometrical arrangement shown in Fig.\,\ref{fig:F1}(a). After non-dimensionalization, Eq.\,\eqref{eq:1}  leads to the governing equation of the classical symmetric bi-stable Duffing oscillator \cite{wang2017harnessing,szemplinska1992cross, mcinnes2008enhanced, carrella2007static, carrella2008static}, 
\begin{equation} \label{eq:2}
\ddot{u}+\gamma \dot{u}-u+\mu u^3=G\cos(\Omega\tau),
\end{equation}
where $\ddot{u} = {d^2 u}/{d \tau^2}$, $\dot{u} = {d u}/{d \tau}$, and the non-dimensional displacement and time are defined as $u=\hat{u}/{l_c}$ and $\tau=t\omega_n$ for the natural frequency $\omega_n=\sqrt{{k_1}/{m}}$ at the linear limit. Other dimensionless parameters are $\gamma={c}/({m\omega_n})$, $\mu={k_3 {l_c}^2}/({m\omega_{n}^2})$, $\Omega={\omega}/{\omega_n}$, and $G={F}/({l_c\omega_{n}^2})$.
For $\mu=1$, Eq.\,\eqref{eq:2} results in  the symmetric energy potential shown in Fig.\,\ref{fig:F1}(b).  If we choose the characteristic length in such a way that $l_c = \sqrt{{k_1}/{k_3}}\;\Rightarrow\; \mu=1$, the potential landscape has double symmetric wells around two stable equilibria at $u_{-1}=-1$ and $u_{+1}=+1$, which are separated by an unstable hilltop equilibrium at $u_{0}=0$ (see \textit{Supplemental Materials}~\cite{SI} for detailed derivation).
The frequency response curve~\cite{wawrzynski2021duffing,ochs2022frequency,brennan2008jump,liu2006comparison,agarwal2018influence,friswell1994accuracy} is a standard tool to describe the bi-stable system dynamics by solving Eq.\,\eqref{eq:2} in the frequency domain, but it does not provide a complete picture. Instead, analyses on the potential well escape provide advantages to investigate the overall dynamics~\cite{szemplinska1995analytical,thompson1989chaotic,lansbury1990incursive,thompson1991indeterminate,stewart1995optimal}. For example, energy criteria~\cite{virgin1994new,gottwald1995routes,virgin1992prediction,gottwald1992experimental,virgin1991note,virgin2000introduction}, forcing phase~\cite{udani2018efficient}, and velocity conditions~\cite{moon1980experiments} offer valuable information. However, due to the uncertainty of the initial conditions~\cite{grebogi1983final,puy2021test} and fractal basin boundaries in the phase space~\cite{moon1985fractal1,nusse1994basins,thompson1987fractal,moon1979magnetoelastic,holmes1979nonlinear,grebogi1987chaos,tarnopolski2013fractal,moon1985fractal,grebogi1982chaotic,moon1980experiments,aguirre2009fractal}, most recent studies are unable to conclusively determine in which potential well (right or left)  the system will reside after the escape. In addition, system parameters such as frequency ($\Omega$), forcing amplitude ($G$), and damping ratio ($\gamma$) also influence the system’s final state. A previous publication~\cite{moon1984fractal} analyzed the forcing amplitude-frequency parameter plane and reported the fractal boundary between the inter-well and intra-well~\cite{wang2017harnessing} behaviors. A fractal dimension of 1.26 was calculated from scarce experimental data. More robust data analyses are necessary to understand the fractal boundary's statistical complexity. Several other studies also reported regions of various shapes~\cite{jan2013phase,bonatto2008chaotic,de2017characterization,prants2014organization} and the Arnold tongue~\cite{paar1998intermingled} in the parameter space of the Duffing system. However, none of them investigated the fractal nature of the region boundaries. 

In this study, we propose numerical criteria addressing the oscillator's categories of behavior between potential wells. To avoid the well-known sensitivity due to the initial condition, we fix the initial conditions at $(u,\dot u)=(-1,0)$ in all simulations. Furthermore, we seek to conduct rigorous numerical simulations to categorize different system behaviors and calculate accurate fractal dimensions in the parameter space.
\begin{figure}[t!]
\centering
\includegraphics[scale=0.63]{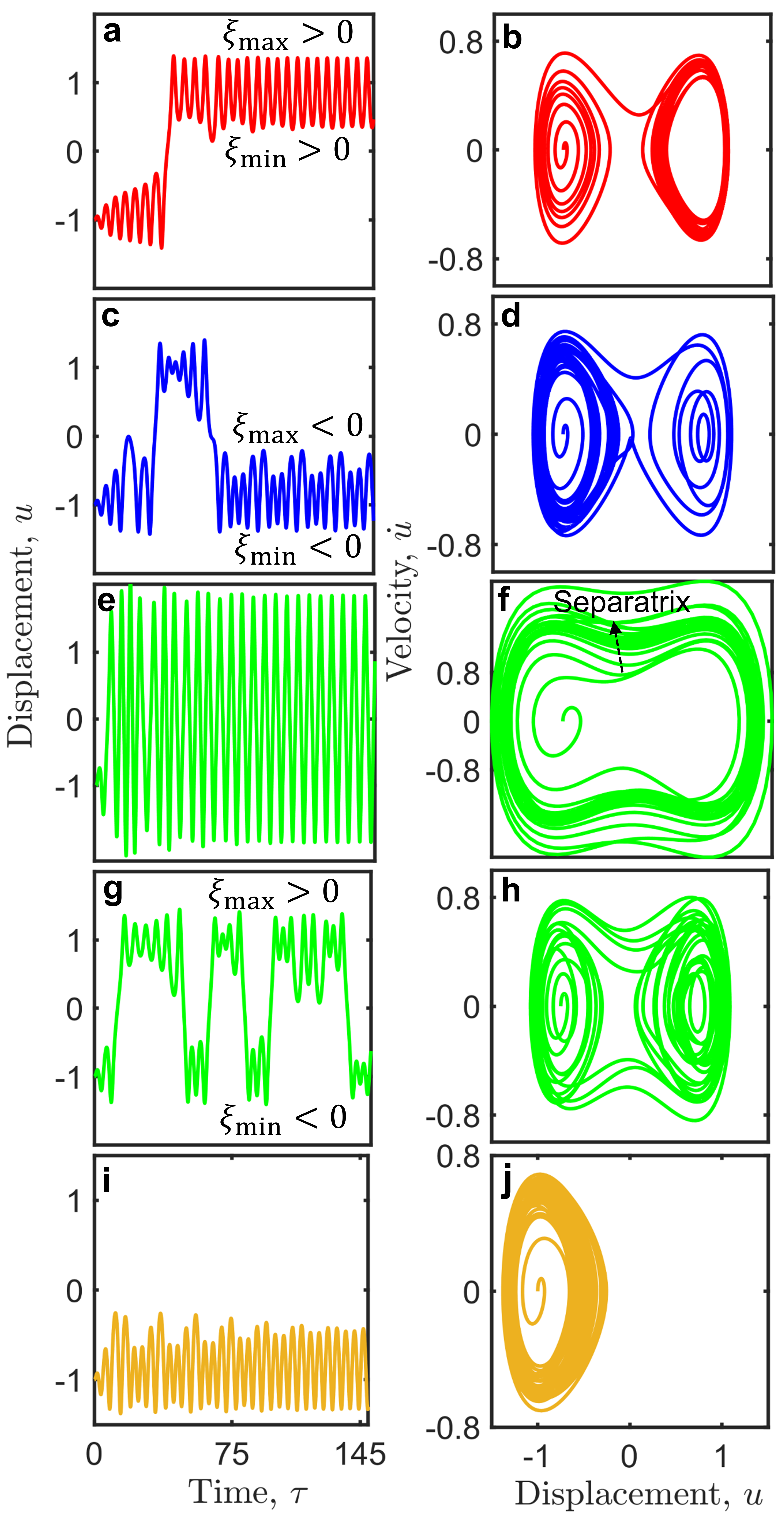}
\caption{\label{fig:F2} 
Categories of time-domain and phase-space of behaviors with damping ratio, $\gamma=0.07$: (a,b) Switching behavior with parameters $(G,\Omega)=(0.100,1.17595)$. (c,d) Reverting behavior with parameters $(G,\Omega)=(0.125,1.209412)$. (e,f) Periodic vacillating behavior with parameters $(G,\Omega)=(0.400,1.10294)$. (g,h) Aperiodic vacillating behavior with parameters $(G,\Omega)=(0.175,1.13291)$. (i,j) Intra-well behavior with parameters $(G,\Omega)=(0.150,1.30506)$}. 
\end{figure} 
We first re-examine the frequency response curve and the parameter space of the bi-stable Duffing oscillator. Importantly, our numerical investigation reveals that the single-degree-of-freedom (SDOF) model based on single-vibration-mode assumption exhibits the fractal pattern very well. Next, we categorize the different steady-state response behaviors inside the parameter space. Finally, we calculate the fractal dimensions of the boundaries that separate the inter-well and intra-well behaviors in the parameter space. The fractal dimension reveals the statistical complexity of the parameter space of the bi-stable Duffing system. Our results indicate that inter-well behavior, traditionally defined as chaotic behavior, can be understood and has potential applications in shape-morphing.

\textit{Frequency response curve and its limitations:} We start with time-domain simulations using the fourth-order Runge-Kutta scheme over a broad range of ($G, \Omega$) for a constant damping ratio of $\gamma = 0.07$ to discern the inter-well and intra-well responses. In Fig.\,\ref{fig:F1}(c), we sweep the forcing frequency $\Omega$ in the range of $0.80\leq\Omega\leq2.0$ at the step size of $0.0047$. We also sweep the forcing amplitude $G$ in the range of $0.100\leq G\leq0.200$ at the step size of $0.025$. In each simulation, we solve Eq.\,\eqref{eq:2} to obtain the time response of 500 forcing cycles. Then we capture the average peak-to-peak steady-state amplitude of the last 50 periods of the time response. In all cases, we start with the initial conditions of $(u,\dot u)=(-1,0)$. 
\begin{figure*}[htb!]
\centering
\includegraphics[scale=0.53]{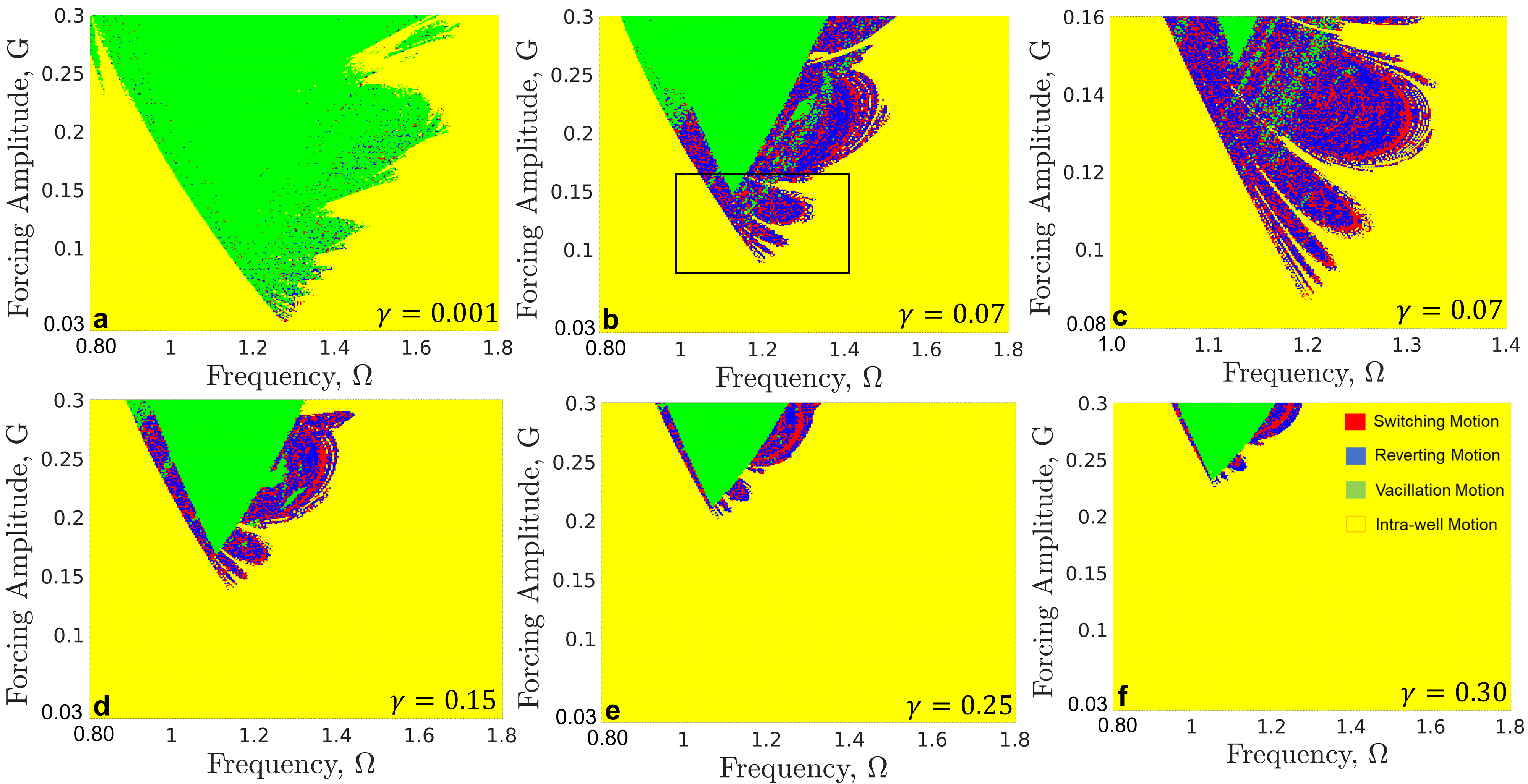}
\caption{\label{fig:F3}
Forcing amplitude-frequency ($G$\,vs.\,$\Omega$) parameter space 
(sampled over a uniform grid of $256\times256$) with: (a) Damping ratio, $\gamma=0.001$ (b) $\gamma=0.07$, (c) Zoom-in of the black-framed portion of Fig.\,\ref{fig:F3}(b),  which shows the fractal nature. (d) $\gamma=0.15$. (e) $\gamma=0.25$. (f) $\gamma=0.30$.}
\end{figure*} 
First, we examine the frequency response curve in Fig.\,\ref{fig:F1}(c). It shows multiple solutions, both inter-well and intra-well, co-existing near $\Omega=1.0$, i.e., the system's resonant frequency at the linear limit. This is the root cause of the hysteresis jump phenomena~\cite{virgin1994new,gottwald1995routes}.  Consequently, the standard frequency response curve cannot provide enough information about potential-well escape criteria. 

We then consider the parameter space of forcing amplitude and driving frequency ($G$\,and\,$\Omega$) and use a $256\times256$ grid to plot the results in Fig.\,\ref{fig:F1}(d). We obtain the time history for 500 periods in each simulation. If the response amplitude, which begins at $u_{-1}=-1$ based on the initial condition, exceeds the hilltop equilibrium at  $u_{0}=0$, then the response exhibits the inter-well behavior. Alternatively, if the response amplitude does not exceed the unstable equilibrium at $u_{0}$ in all 500 periods, we recognize this response as the intra-well behavior. The two different parameter regions are illustrated in Fig.\,\ref{fig:F1}(d). As examples, Figs.\,\ref{fig:F1}(e) and (f) show the time and phase space responses of the inter-well (magenta curves) and intra-well (yellow curves) behaviors. 
We observe that the boundary between the inter-well and intra-well zones in Fig.\,\ref{fig:F1}(d) takes an intricate shape. Near the boundary, the oscillator's behavior is highly sensitive to a subtle change in the amplitude and frequency of the excitation.  
This shows that the single-mode single-degree-of-freedom (SDOF) model can capture fractal patterns in the parameter space of a bi-stable system, contrary to the argument that higher-order vibration modes with more degrees of freedom are required~\cite{moon1984fractal}.
\begin{figure*}[t!]
\centering
\includegraphics[scale=0.55]{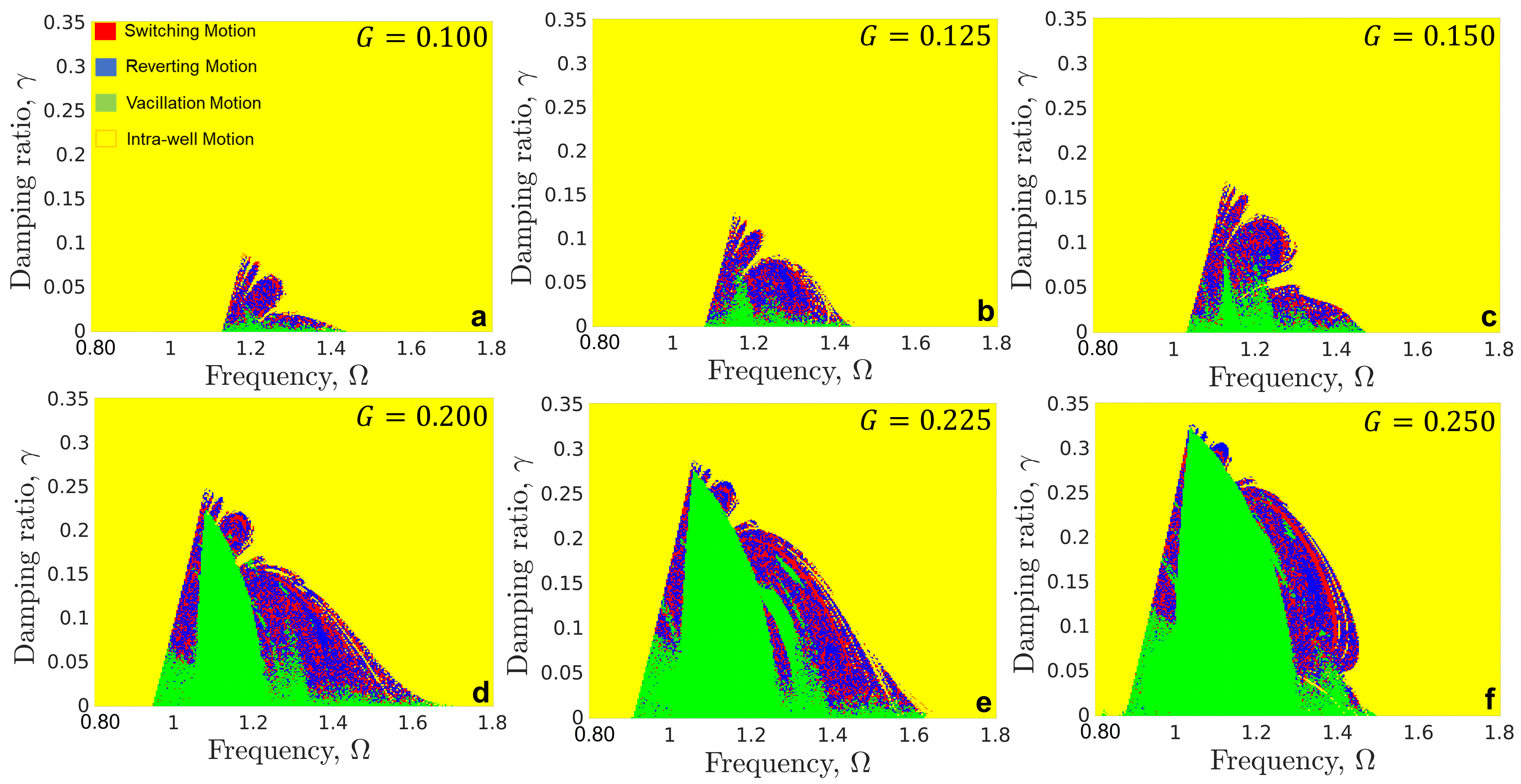}
\caption{\label{fig:F4}  
Damping ratio-forcing frequency ($\gamma$\,vs.\,$\Omega$) parameter space (sampled over a uniform grid of $256\times256$) with: (a) Forcing amplitude, $G=0.100$. (b) $G=0.125$. (c) $G=0.150$. (d) $G=0.200$. (e) $G=0.225$. (f) $G=0.250$.}
\end{figure*}

\textit{Categories of steady-state behaviors in the parameter space:} Fig.\,\ref{fig:F1} shows two behaviors: intra-well oscillation near one stable equilibrium; and inter-well between two stable equilibria. However, a thorough investigation of the dynamics of a double-well potential system with harmonic excitation uncovers opportunities for unique applications, as it exhibits four possible types of behavior:
(i) Switching - inter-well behavior where the oscillation begins in one potential well, and after the transient stage, reaches its steady-state in the other potential well, (ii) Reverting - inter-well behavior where the oscillation begins in one potential well, reaches the other potential well during the transient stage, and eventually returns to its original potential well in the steady state; (iii) Vacillating - inter-well behavior where the oscillation keeps moving between the two wells, and it never settles into either one; and (iv) Intra-well - steady state oscillation in one potential well only.

To quantitatively distinguish these categories of behavior using numerical results from time-domain simulations, we first check whether the system displacement amplitude, which begins at $u_{-1}$ based on the initial condition, exceeds the hilltop equilibrium at $u_{0}$. Then, we recognize the last 50 cycles as the steady-state output of the total of 500 cycles in the simulations. Based on this numerical assumption, we define $\xi_\textrm{max} = \max(u(t)| t \in \{\text{last 50 cycles}\})$ and $\xi_\textrm{min} = \min(u(t)| t \in \{\text{last 50 cycles}\})$. As illustrated in Fig.\,\ref{fig:F2}(a) - (Note that a much smaller number of cycles are shown here for illustration purposes), if $\xi_{\textrm{max}} > 0$ and $\xi_{\textrm{min}} > 0$, we categorize this behavior as switching. In Fig.\,\ref{fig:F2}(b), the phase space plot of the same simulation illustrates the growth of the oscillator's limit cycle oscillations (LCOs)~\cite{udani2017sustaining} from $u_{-1}$, and it eventually reaches a steady-state trajectory near $u_{+1}$. This picture resembles the stable LCOs in many studies~\cite{udani2017sustaining,udani2018efficient,strogatz2018nonlinear}. Thus, up to the numerical precision, we can conclude that the oscillator stabilizes in the second stable state~\cite{strogatz2018nonlinear} and categorize this as the switching behavior. In Fig.\,\ref{fig:F2}(c), the oscillator switches from the $u_{-1}$ potential well to the $u_{+1}$ one during the transient stage. For the steady-state, we observe $\xi_{\textrm{max}}<0$ and $\xi_{\textrm{min}}<0$, meaning it reverts back to its initial well at $u_{-1}$ in the end. We categorize this behavior as reverting. Correspondingly, we see a stable limit cycle around the stable equilibrium at $u_{-1}$ in Fig.\,\ref{fig:F2}(d). Next, in Figs.\,\ref{fig:F2}(e) and (g), the oscillator's inter-well behavior exhibits persistent shifting between $u_{-1}$ to $u_{+1}$. The observation of $\xi_{\textrm{max}}>0$ and $\xi_{\textrm{min}}<0$ in the steady-state implies that it continues to oscillate between both potential wells without settling into either one, so we can categorize these behaviors as vacillating. We also analyze the phase space behavior to distinguish between periodic and aperiodic vacillating behaviors. Fig.\,\ref{fig:F2}(f) illustrates periodic vacillating where the phase space trajectory follows a high-energy orbit motion outside of the separatrix~\cite{udani2017sustaining}, while Fig.\,\ref{fig:F2}(h) shows low-energy orbits around either one or two potential wells inside the separatrix. Finally, Fig.\,\ref{fig:F2}(i) displays the intra-well behavior, which remains in the vicinity of the stable equilibrium at $u_{-1}$ without reaching the unstable equilibrium at $u_{0}$. Fig.\,\ref{fig:F2}(j) shows the stable limit cycle around $u_{-1}$ corresponding to the intra-well confinement.
\begin{figure*}[t!]
\centering
\includegraphics[scale=0.38]{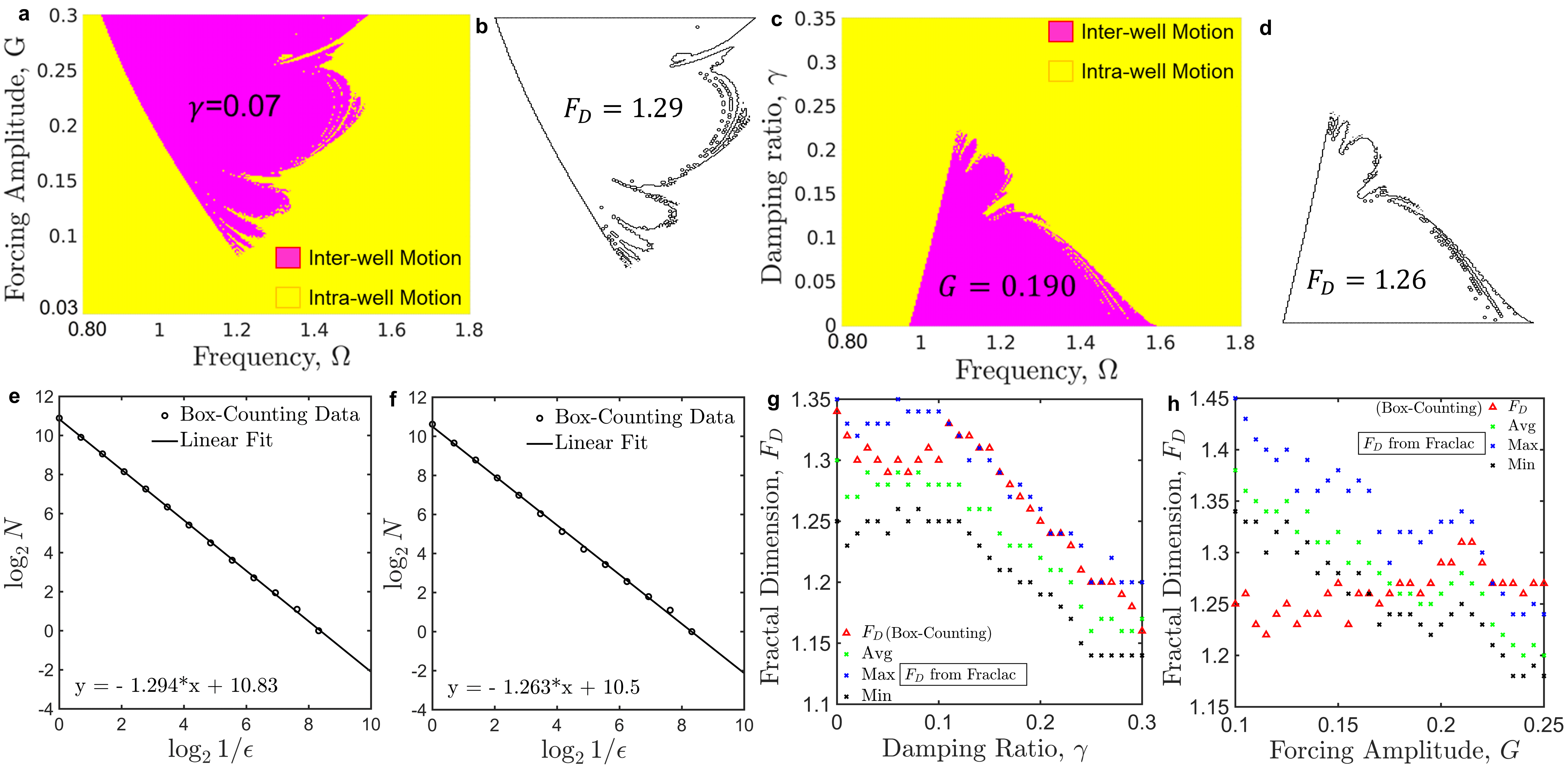}
\caption{\label{fig:F5}
Fractal dimension of the boundaries in forcing amplitude-frequency ($G$\,vs.\,$\Omega$) and damping ratio-forcing frequency ($\gamma$\,vs.\,$\Omega$) parameter space: (a) An example with the damping ratio of $\gamma=0.07$ (b) Fractal boundary between inter-well and intra-well regions. (c) An example with the forcing amplitude of $G=0.190$ (d) Fractal boundary between inter-well and intra-well regions. (e) and (f) show the linear regression of the box-counting algorithm data for (b) and (d) using the least square method to calculate the Hausdorff fractal dimension. (g) and (h) show the downward trend of the Hausdorff fractal dimension of the boundaries in parameter space as the damping ratio and forcing amplitude increase (see \textit{Supplemental Materials}~\cite{SI} for more results).}
\end{figure*}

\textit{Fractal patterns in the parameter space:} Using the four categories of behaviors described in the last section, we update our understanding of the parameter space. For each set of ($G$, $\Omega$), we run time-domain simulations in the parameter ranges of $0.80\leq\Omega\leq1.8$ and $0.03\leq G\leq0.30$ with different damping ratios. Figs.\,\ref{fig:F3}(a)-(f) show the results for $\gamma=$ 0.001, 0.07, 0.15, 0.25, and 0.30, respectively (see \textit{Supplemental Materials}~\cite{SI} for more results with other damping ratios). We categorize the numerical steady state of each simulation either as switching, reverting, vacillating, or intra-well behavior. We represent them as red, blue, green, and yellow data points in Fig.\,\ref{fig:F3}, respectively. The minimally required forcing amplitude $G_\textrm{min}$ for inter-well behavior rises as $\gamma$ increases. For example, Fig.\,\ref{fig:F3}(a) indicates the minimum amplitude required for inter-well behavior is $G=0.04$. In contrast, in Figs.\,\ref{fig:F3}(b)\,and\,\ref{fig:F3}(c), it is about $G=0.09$. In all cases, we see the well-known V-shaped ``Arnold's tongue", whose bottom tip represents $G_\textrm{min}$ for inter-well behavior. As damping increases, the tip of Arnold’s tongue moves up and left towards $\Omega=1$, which is the system's resonant frequency at the linear limit. Fig.\,\ref{fig:F3}(a), in particular, shows that the vacillating behavior dominates among all inter-well behaviors in the low damping limit. This means that low dissipation increases the likelihood of vacillating behavior. 

In general, the results in Figs.\,\ref{fig:F3} suggest that, compared to the quasi-static actuation~\cite{shan2015multistable}, utilizing harmonic excitation can significantly reduce the force amplitude necessary to switch between stable states. For the same system with static loading, the force required for switching between stable states is $0.38$ (see \textit{Supplemental Materials}~\cite{SI}). In contrast, Figs.\,\ref{fig:F3} reveals that inter-well behavior can be achieved with much lower actuation peak force with dynamic loading. 
These results highlight the importance of optimizing the $(G,\Omega,\gamma)$ parameters in achieving the inter-well behavior of the system. Furthermore, incorporating feedback control strategy~\cite{sadeghi2020dynamic} can further improve controlled switching between stable states, leading to a much more efficient actuation mechanism than quasi-static actuation.

To better understand the system behaviors, in Fig.\,\ref{fig:F3}(c), we zoom in on the black-framed region in Fig.\,\ref{fig:F3}(b). 
Fig.\,\ref{fig:F3}(c) shows intricate feather-shaped self-similar patterns and strongly suggests the existence of fractals in the parameter space of the bi-stable Duffing oscillator. Motivated by the fractal pattern in the forcing amplitude-frequency ($G$\,vs.\,$\Omega$) parameter plane, we also investigate the damping ratio-forcing frequency ($\gamma$\,vs.\,$\Omega$) parameter plane. Fig.\,\ref{fig:F4} shows the results in the parameter range $0.80\leq \Omega \leq 1.8$ and $0.00 \leq \gamma \leq 0.35$ for $G=$ 0.1, 0.125, 0.15, 0.2, 0.225, and 0.25, respectively (see \textit{Supplemental Materials}~\cite{SI} for more results with other forcing amplitudes).
The patterns look similar to those shown in Fig.\,\ref{fig:F3}, but they are flipped upside down with the top tip indicating the maximally allowed damping for inter-well behavior. Like Fig.\,\ref{fig:F3}, with increasing forcing amplitude $G$, the ``Arnold's tongue" tip moves up and also left towards  $\Omega=1$, the resonant frequency of the system at the linear limit. Feather-shaped fractal patterns with self-similarity also appear. The results show that higher forcing amplitude and lower dissipation lead to a higher likelihood of vacillating behavior in the steady state.

\textit{Fractal dimensions of boundaries in parameter space:} To understand the system behavior's complexity, we calculate the Hausdorff fractal dimension~\cite{moon2008chaotic}, $F_\textrm{D}$. As shown in Figs.\,\ref{fig:F5}(a) and \ref{fig:F5}(c), we first numerically categorize the inter-well and intra-well behavior types. Then, the boundary between the inter-well and intra-well categories can be extracted. After that, as illustrated in Figs.\,\ref{fig:F5}(b) and \ref{fig:F5}(d), we implement the box-counting algorithm~\cite{wu2020effective,Alceau} to calculate the Hausdorff fractal dimension of extracted boundary curves in each parameter space plot. We use square-shaped boxes of variable size $\epsilon$ to cover the boundary curves [e.g., black lines in Figs.\,\ref{fig:F5}(b) and (d)], and count the total number of boxes, $N(\epsilon)$, needed to cover all boundaries. As $\epsilon \rightarrow 0$, $N(\epsilon) \rightarrow \infty$, and we obtain $F_\textrm{D}$ as,
\begin{equation}\label{eq:3}
F_\textrm{D} = \Big| \lim_{\epsilon\to 0} \big({\log (N)} / {\log(\frac{1}{\epsilon})}\big) \Big|.
\end{equation}
Finally, we use the least-square method to perform linear regression on the $\log(N)$ vs. $ \log(\frac{1}{\epsilon})$ plot to determine $F_\textrm{D}$. Detailed procedures are further explained in the \textit{Supplemental Materials}~\cite{SI}. For example, the linear regressions shown in Figs.\,\ref{fig:F5}(e) and \ref{fig:F5}(f) generate $F_\textrm{D}=1.29$ and $F_\textrm{D}=1.26$ for the parameter space boundaries shown in Figs.\,\ref{fig:F5}(b) and \ref{fig:F5}(d), respectively. To verify our implementation, we also use the software \textsc{Fraclac/ImageJ}~\cite{Fraclac} developed by the National Institute of Health (NIH). It determines the fractal dimensions of Figs.\,\ref{fig:F5}(b) and \ref{fig:F5}(d) as $1.2813$ and $1.2545$, respectively, which closely match our numerical predictions of $F_\textrm{D}$ from Figs.\,\ref{fig:F5}(e) and \ref{fig:F5}(f). Following the above procedure, we calculate the fractal dimensions of boundaries in 31 forcing amplitude-frequency plots and 31 damping ratio-forcing frequency plots of the parameter space (see \textit{Supplemental Materials}~\cite{SI}) for $0.001 \leq \gamma \leq 0.30$ and $0.10\leq G \leq 0.25$, respectively. Using \textsc{Fraclac/ImageJ}~\cite{Fraclac}, we also calculate the average, maximum, and minimum fractal dimensions in all cases. We summarize all results into Fig.\,\ref{fig:F5}(g) for different damping and Fig.\,\ref{fig:F5}(h) for different forcing amplitude, respectively. Our results indicate that the Hausdorff fractal dimension of the boundaries between behavior categories decreases with an increase in the damping ratio and forcing amplitude.\\
\textit{Conclusion:} We show that four types of system behaviors (switching, reverting, vacillating, and intra-well) exist as different regions in the parameter space of a dissipative bi-stable Duffing oscillator. Our numerical results reveal the fractal nature of the boundaries between the inter-well and intra-well regions, and the box-counting algorithm leads to the characterization of the Hausdorff fractal dimension of these boundaries. This study exclusively focuses on the fractal patterns in the parameter space, often neglected compared to fractal patterns in the phase space. These findings may have important implications for the design of advanced mechanical metamaterials for shape-morphing applications. 

We acknowledge the support from the National Institutes of Health (NIH): Project No.\,1R01EB032959-01. Start-up funds from the Department of Mechanical Engineering at the Univ.\,of\,Utah also supported this work. The support and resources from the Center for High-Performance Computing at Univ.\,of\,Utah are gratefully acknowledged. The authors thank Prof.\,Lawrence Virgin (Duke Univ.), Prof.\,Suyi Li (Virginia Tech.), and  Prof.\,Andres Arrieta (Purdue Univ.) for inspirational discussions.

\normalem
\bibliography{ref_NoTitle}      

\end{document}


\title{Supporting Information for \\
\emph{Rolling Waves with Non-Paraxial Phonon Spins}}

\author{Peng Zhang}
\affiliation{Department of Mechanical Engineering, University of Utah, Salt Lake City, UT 84112, USA}

\author{Christian Kern}
\affiliation{Department of Mathematics, University of Utah, Salt Lake City, UT 84112, USA}

\author{Sijie Sun}
\affiliation{Harvard John A. Paulson School of Engineering and Applied Science, Harvard University, Cambridge, MA 02138, USA}

\author{David A. Weitz}
\affiliation{Harvard John A. Paulson School of Engineering and Applied Science, Harvard University, Cambridge, MA 02138, USA}

\author{Pai Wang}%
\affiliation{Department of Mechanical Engineering, University of Utah, Salt Lake City, UT 84112, USA}
\thanks{pai.wang@utah.edu}

\maketitle